\documentstyle[12pt]{article}
\title{\bf Dynamical conformal transformation \\and \\classical Euclidean wormholes }
\author{F. Darabi \thanks{f.darabi@azaruniv.edu} \\
{\small Department of Physics, Azarbaijan University of Tarbiat
Moallem, 53714-161 Tabriz, Iran}\\
{\small Research Institute for Astronomy and Astrophysics of
Maragha (RIAAM), 55134-441 Maragha, Iran }}
\begin{document}
\maketitle
\begin{abstract}
We investigate the necessary condition for the existence of classical Euclidean wormholes in a conformally non-invariant gravitational model minimally coupled
to an scalar field. It is shown that while the original Ricci tensor with positive eigenvalues does not allow the Euclidean wormholes to occur, under dynamical conformal transformations the Ricci tensor, with respect to the original metric, is dynamically coupled with the conformal field and
its eigenvalues may become negative allowing the Euclidean wormholes to occur. Therefore, it is conjectured that dynamical conformal transformations may provide us with {\it effective} forms of matter sources leading to Euclidean wormholes in conformally non-invariant systems.
\end{abstract}

\newpage

\section{Introduction}

Classical wormholes are usually considered as Euclidean metrics
that consist of two asymptotically flat regions connected by a
narrow throat (handle). Wormholes have been studied mainly as
instantons, namely solutions of the classical Euclidean field
equations. In general, such wormholes can represent quantum
tunneling between different topologies. They are possibly useful
in understanding black hole evaporation \cite{Haw}; in allowing
nonlocal connections that could determine fundamental constants;
and in vanishing the cosmological constant $\Lambda$ \cite{Col1}-\cite{Col3}.
They are even considered as an alternative to the Higgs mechanism
\cite{Mig}. Therefore, an open and interesting problem is whether Euclidean wormholes can occur for fairly general matter sources.
The non existence of instantons, for general matter sources, not
only makes it difficult to believe  that wormholes are the mechanism
for black hole evaporation but also casts doubt on whether wormholes
are the reason why the cosmological constant is zero. Due to limited
known classical wormhole solutions, Hawking and page advocated a
different approach in which wormholes were regarded, not as
solutions of the classical Euclidean field equations, but as
solutions of the quantum mechanical Wheeler-DeWitt equation
\cite{HP}. These wave functions have to obey certain boundary
conditions in order that they represent wormholes. The boundary
conditions seem to be that the wave function is exponentially damped
for large tree geometries, and is regular in some suitable way when
the tree-geometry collapses to zero. 

The reason why classical wormholes may exist is related to the
implication of a theorem of Cheeger and Glommol \cite{Chee} which
states that a {\it necessary} (but not sufficient) condition for a classical wormhole to exist is that the eigenvalues of the Ricci tensor be negative {\it somewhere} on the manifold. Unfortunately, there exists certain special kinds of matter for which the above necessary condition is satisfied. 
For example, the energy-momentum tensors of an axion field and of a conformal
scalar field are such that, when coupled to gravity, the Ricci
tensor has negative eigenvalues. However, for pure gravity or a (real) minimally coupled scalar field it is easy to show that the Ricci tensor can never be negative. It is shown that the (pure imaginary) minimally coupled scalar field, conformally coupled scalar field, a third rank antisymmetric tensor field and some special kinds of matter sources have wormhole solutions \cite{Gid1}-\cite{Gid5}.

The theory of conformal transformations, on the other hand, has been playing a particularly important role in the investigation
of gravitational models since Weyl, who introduced the notion of conformal rescaling of the metric tensor. Afterwards, it was promoted to the conformal transformations in scalar tensor theories, in which another transformation on the scalar field was required to represent the conformal invariance in modern gravitational models. Conformal invariance implies that the gravitational theory is invariant under local changes of units of length and time. These local transformations relate different unit systems or conformal frames via {\it arbitrary} space time dependent conformal factors, and these unit systems are dynamically distinct. Therefore, the introduction of conformal transformations into the gravitational theory gives rise to a new dynamical field so called {\it conformal field}. 

In the present paper we investigate the role of dynamical conformal
transformations in providing the possibility for the existence of classical Euclidean wormholes in a minimally coupled real scalar field gravitational model. In more precise words, we shall show that a conformal transformation in a conformally non-invariant system usually changes the field equation such that the originally positive eigenvalues of the Ricci tensor may change to negative ones due to the presence of dynamical
conformal field. In fact, it seems reasonable that (especially in small scales) the fluctuation of conformal factor as a dynamical field may lead to the appearance of (microscopic) Euclidean wormholes. Beside the Hawking-Page conjecture, this may provide us with the possibility of obtaining more Euclidean wormholes for {\it effective} matter sources resulting from the dynamics
of conformal field.

\section{No wormholes with minimally coupled real scalar field}

It is well known that classical Euclidean wormholes can occur if the Ricci
tensor has negative eigenvalues somewhere on the manifold. Actually, this
is necessary but not sufficient condition for their existence
and is related to the implication of a theorem of Cheeger and Glommol \cite{Chee}.
For example, the energy-momentum tensors of an axion field and of a conformal
scalar field are such that, when coupled to gravity, the Ricci
tensor has negative eigenvalues. However, for a minimally coupled
scalar field with the lagrangian density
\begin{equation}\label{1''}
{\cal L}=\frac{1}{2}\nabla_{\mu}\phi \nabla^{\mu}\phi-V(\phi),
\end{equation}
the energy-momentum tensor and its trace are given
\begin{equation}\label{2''}
T_{\mu \nu}=\nabla_{\mu}\phi \nabla_{\nu}\phi-\frac{1}{2}g_{\mu
\nu}\nabla_{\lambda}\phi\nabla^{\lambda}\phi-g_{\mu \nu}V(\phi),
\end{equation}
\begin{equation}\label{3''}
T=-\nabla_{\mu}\phi \nabla^{\mu}\phi-4V(\phi).
\end{equation}
The Einstein equations 
\begin{equation}\label{4''}
R_{\mu \nu}=T_{\mu \nu}-\frac{1}{2}g_{\mu \nu}T,
\end{equation}
give
\begin{equation}\label{5''}
R_{\mu \nu}=\nabla_{\mu}\phi \nabla_{\nu}\phi+ g_{\mu \nu}V(\phi),
\end{equation}
which shows $R_{\mu \nu}$ can never be negative in Euclidean space $g_{\mu \nu}=(++++)$ for $V(\phi)\geq 0$ unless we consider
a purely imaginary scalar field, i.e., let $\phi \rightarrow i\phi$ \cite{Coule}. 
\section{Possible wormholes with minimally coupled real scalar field subject
to dynamical conformal transformations}

The theory of conformal transformation has been playing a particularly important role in the investigation of gravitational models since Weyl, who introduced the notion of conformal rescaling of the metric tensor. Afterwards, it was promoted to the conformal transformations in scalar tensor theories, in which another transformation on the scalar field was required to represent the conformal invariance in modern gravitational models. Conformal invariance implies that the gravitational theory is invariant under local changes of units of length and time. These local transformations relate different unit systems or conformal frames via arbitrary space time dependent conformal factors, and these unit systems are dynamically distinct. Therefore, the introduction of conformal transformations into the gravitational theory gives rise to a new dynamical field so called {\it conformal field}. 
Let us now consider the following conformal transformations 
\begin{equation}\label{8''}
g_{\mu \nu} \rightarrow \bar{g}_{\mu \nu}=\Omega^2g_{\mu \nu},\:\:\:
\phi\rightarrow \bar{\phi}=\Omega^{-1} \phi,
\end{equation}
where $\Omega(x)$ is a continuous, non-vanishing, finite, real arbitrary
function. For such a transformation of the metric we obtain the following
transformation for the Ricci tensor \cite{Wald}
\begin{eqnarray}\label{7''}
R_{\mu \nu}\rightarrow \bar{R}_{\mu \nu} &=& 
R_{\mu \nu}-(n-2)\nabla_{\mu}\nabla_{\nu}\ln
\Omega-g_{\mu \nu}g^{\rho \sigma}\nabla_{\rho}\nabla_{\sigma}\ln
\Omega \\ \nonumber
&+& (n-2)(\nabla_{\mu}\ln
\Omega)(\nabla_{\nu}\ln
\Omega)-(n-2)g_{\mu \nu}g^{\rho \sigma}(\nabla_{\rho}\ln
\Omega)(\nabla_{\sigma}\ln
\Omega),
\end{eqnarray}
where $n$ is the dimension of the manifold. Now, we shall assume the conformal
factor to be a new dynamical degree of freedom and let the transformed action to involve a kinetic term for $\Omega$ 
\begin{equation}\label{1'''}
{\cal {L'}}=\frac{1}{2}\bar{\nabla}_{\mu}\bar{\phi} \bar{\nabla}^{\mu}\bar{\phi}-V(\bar{\phi})+
\frac{1}{2}{\nabla}_{\mu}\Omega {\nabla}^{\mu}\Omega.
\end{equation}
Note that the kinetic term for the conformally transformed scalar field $\bar{\phi}$ is properly defined in terms of the conformally transformed metric 
$\bar{g}_{\mu \nu}$. However, since the conformal field $\Omega$ itself does not transform and is just defined with respect to $g_{\mu \nu}$ to yield $\bar{g}_{\mu \nu}$, we have considered the kinetic term for $\Omega$ with respect to the metric $g_{\mu \nu}$.  Actually, one may also define the kinetic term for the conformal field $\Omega$ by  $\bar{g}_{\mu \nu}$, but then the corresponding field in terms of the metric $g_{\mu \nu}$ will be $\ln \Omega$ which is not so reasonable. Nevertheless, Using $g_{\mu \nu}$ or $\bar{g}_{\mu \nu}$ does not change the main result in this paper,
as we shall see bellow.

If we now impose (\ref{8''}) into the system of minimally coupled real scalar
and conformal fields we obtain the new Einstein field equation
\begin{equation}\label{9''}
\bar{G}_{\mu \nu}=\bar{T}_{\mu \nu},
\end{equation}
or
\begin{equation}\label{10''}
\bar{R}_{\mu \nu}=\bar{T}_{\mu \nu}-\frac{1}{2}\bar{g}_{\mu \nu} \bar{T}.
\end{equation}
Using Eq.(\ref{7''}) for $n=4$ we may rewrite Eq.(\ref{10''}) as follows
\begin{eqnarray}\label{11''}
R_{\mu \nu}&=& \bar{T}_{\mu \nu}-\frac{1}{2}\bar{g}_{\mu \nu} \bar{T}\\ \nonumber
&+& 2\nabla_{\mu}\nabla_{\nu}\ln
\Omega 
+g_{\mu \nu}g^{\rho \sigma}\nabla_{\rho}\nabla_{\sigma}\ln
\Omega \\ \nonumber
&-& 2(\nabla_{\mu}\ln
\Omega)(\nabla_{\nu}\ln
\Omega)+2g_{\mu \nu}g^{\rho \sigma}(\nabla_{\rho}\ln
\Omega)(\nabla_{\sigma}\ln
\Omega),
\end{eqnarray}
where the transformed energy-momentum tensor and its trace are given by
\begin{eqnarray}\label{12''}
\bar{T}_{\mu \nu}&=&\Omega^{-2}[\nabla_{\mu}\phi \nabla_{\nu}\phi-\frac{1}{2}g_{\mu
\nu}\nabla_{\lambda}\phi\nabla^{\lambda}\phi]\\ \nonumber
&+&\Omega^{-1}[\phi\nabla_{\mu}\phi\nabla_{\nu}\Omega^{-1}+\phi\nabla_{\mu}\Omega^{-1}\nabla_{\nu}\phi-
g_{\mu
\nu}\phi\nabla_{\lambda}\phi\nabla^{\lambda}\Omega^{-1}]\\ \nonumber
&+&\phi^2[\nabla_{\mu}\Omega^{-1} \nabla_{\nu}\Omega^{-1}-\frac{1}{2} g_{\mu
\nu}\nabla_{\lambda}\Omega^{-1}\nabla^{\lambda}\Omega^{-1}]-g_{\mu
\nu}\Omega^{2}V(\Omega^{-1}\phi)+\nabla_{\mu}\Omega \nabla_{\nu}\Omega-\frac{1}{2}g_{\mu
\nu}\nabla_{\lambda}\Omega\nabla^{\lambda}\Omega,
\end{eqnarray}
and
\begin{eqnarray}\label{13''}
\bar{T}=&-&\Omega^{-4}\nabla_{\lambda}\phi \nabla^{\lambda}\phi-\Omega^{-2}\phi^2\nabla_{\lambda}\Omega^{-1} \nabla^{\lambda}\Omega^{-1}\\ \nonumber
&-&2\phi \Omega^{-3}\nabla_{\lambda}\phi \nabla^{\lambda}\Omega^{-1}
-4V(\Omega^{-1}\phi)-\Omega^{-2}\nabla_{\mu}\Omega \nabla^{\mu}\Omega.
\end{eqnarray}
Putting (\ref{12''}) and (\ref{13''}) into (\ref{11''}) we obtain the field
equation
\begin{eqnarray}\label{14''}
R_{\mu \nu}= \Omega^{-2}\nabla_{\mu}\phi \nabla_{\nu}\phi&+&g_{\mu
\nu}\Omega^{2}V(\Omega^{-1}\phi)+
\phi^2\nabla_{\mu}\Omega^{-1} \nabla_{\nu}\Omega^{-1}+\nabla_{\mu}\Omega \nabla_{\nu}\Omega\\ \nonumber
&+& 2\nabla_{\mu}\nabla_{\nu}\ln
\Omega 
+g_{\mu \nu}g^{\rho \sigma}\nabla_{\rho}\nabla_{\sigma}\ln
\Omega 
- 2(\nabla_{\mu}\ln
\Omega)(\nabla_{\nu}\ln
\Omega)\\ \nonumber
&+&2g_{\mu \nu}g^{\rho \sigma}(\nabla_{\rho}\ln
\Omega)(\nabla_{\sigma}\ln
\Omega)+\Omega^{-1}\phi[\nabla_{\mu}\phi\nabla_{\nu}\Omega^{-1}+\nabla_{\mu}\Omega^{-1}\nabla_{\nu}\phi],
\end{eqnarray}
which reduces to the original Ricci tensor (\ref{5''}) for $\Omega=1$. 

\section{Discussion}

Suppose we have a conformally invariant gravitational action where the Ricci
tensor has not negative eigenvalues, so there is no Euclidean wormhole solution.
Under a conformal transformation it is obvious that the same situation holds in the new conformal frame due to the conformal invariance of the
action which results in no Euclidean wormhole solution. The action of real scalar field, corresponding to (\ref{1''}), coupled minimally with gravity however is not generally conformal invariant, so we expect nontrivial effects
in comparison with conformal invariant systems. Note that a conformal transformation is different from a scale transformation. A scale transformation is just a re-scaling of metric by a space time dependent conformal factor, and all lengths are assumed to remain unchanged (like co-moving coordinates in cosmology). The scale transformation is not a {\it unit} transformation; it is just a dynamical re-scaling (enlargement or contraction) of a system. But, a conformal transformation is viewed as stretching all lengths by a space time dependent conformal factor, namely a {\it unit} transformation.
A dynamical conformal transformation is therefore responsible for dynamical transformation of any dimensional coupling in the theory. 

To the extent we
are dealing with a scale transformation it is obvious that this transformation
is unable to change the topology of a system in a way that an Euclidean solution with wormhole topology emerges from a situation where no Euclidean wormhole
topology is there. This is simply because scale transformation can not change
the topology of the manifold. But, the story in the case of conformal transformation is completely different. It may affect drastically the matter source especially for a conformally non-invariant system. Therefore, new {\it effective} forms of matter sources can emerge resulting in the possibility for occurrence of Euclidean wormholes.
In fact, under a conformal transformation some extra terms involving the dynamics of the conformal field $\Omega$ appear, in a nontrivial way, in the field equations. This is what we have obtained in (\ref{14''}): the left hand side is the Ricci tensor corresponding to the original metric $g_{\mu \nu}$ minimally coupled with the scalar field $\phi$ but non-minimally coupled with the conformal field $\Omega$.

The important result is that while the original Ricci tensor, before conformal
transformation, does not allow the appearance of Euclidean wormholes for minimally coupled real scalar field, the new Ricci tensor after conformal transformation
may allow Euclidean wormholes for the 
new system obtained effectively by inclusion of dynamical conformal field
$\Omega$ playing the role of {\it effective} matter field.
This is because the eigenvalues of the Ricci tensor may become
negative due to the dynamical contributions of conformal field $\Omega$. The converse is also true: If the original conformal non-invariant 
system could provide us with Euclidean wormholes, the conformally transformed
system may not allow for the appearance of Euclidean wormholes. Since in the general theory the existence or non-existence of Euclidean wormholes crucially depends on the kind of matter coupled to the metric, then we may conclude that the conformal transformation effectively introduces the dynamical conformal field $\Omega$ as the effective matter source which determines the existence or non-existence of the wormholes. This scenario is mostly plausible at small
scales where the metric fluctuation may provide us with the
dynamical conformal factor leading to the microscopic wormholes.

\section*{Acknowledgment}
This work has been supported financially by Research Institute
for Astronomy and Astrophysics of Maragha.

\end{document}